\documentclass{article}
\usepackage[utf8]{inputenc}
\usepackage{amsmath}
\usepackage{amssymb}
\usepackage{graphicx}
\usepackage{color}

\title{A bound on superconducting $T_c$'s}
\author{I. Esterlis, S. A. Kivelson, and D. J. Scalapino }
\date{\today}

\begin{document}

\maketitle

It is notoriously difficult to make quantitative theoretical predictions of the superconducting $T_c$, either from first-principles  or even from a knowledge of normal state properties.  Ultimately, this reflects the fact that the energy scales involved in the superconducting state are extremely small in natural units, and that $T_c$ depends exponentially on a subtle interplay between different interactions so that small uncertainties in microscopic processes can lead to order 1 effects on $T_c$.  However, in some circumstances, it may be possible to determine (approximate) bounds on $T_c$.  Here, we propose  such a bound for the conventional phonon-mediated mechanism of pairing with strongly retarded interactions, {\it i.e.} in the case in which $\hbar\bar \omega \ll E_F$ where $\bar \omega$ is an appropriate characteristic phonon frequency and $E_F$ is the Fermi energy.  Specifically, drawing on both empirical results (shown in Figure \ref{fig:Holstein_bound} below) and recent results\cite{PhysRevB.97.140501} of determinant quantum Monte Carlo (DQMC) studies of the paradigmatic Holstein model, we propose that
\begin{equation}
    k_B T_c 
    \leq A_{max} \ \hbar \bar \omega
\end{equation}
where 
$A_{max}$ is a dimensionless number of order one that we estimate to be 
\begin{equation}
    A_{max} \approx 1/10.
\end{equation}

\section{Theoretical analysis}
In general, $T_c$ is a function of the phonon spectrum, the electron spectrum, the electron-phonon coupling and the electron-electron interactions.  Given that the bare interactions between electrons is repulsive, it is something of a miracle that the effective interactions at low energies can be effectively attractive, {\it i.e.} produce pairing.  In conventional superconductors, this miracle is a consequence of the retarded character of the phonon induced interactions, $\hbar\bar \omega \ll E_F$.  (In unconventional superconductors, it is rather the strong $\vec k$ dependence of the screened but still largely repulsive effective interactions that allows pairing to occur.)  Since we are focusing here on the case of conventional superconductors, it is reasonable to write a general expression for $T_c$ as the product of a dimensional factor $\hbar \bar \omega/k_B$  times a dimensionless function of the various dimensionless parameters that characterize the particular system in question
\begin{equation}
    k_BT_c = \hbar \bar \omega \ A(w,\lambda,\mu, \ldots)
    \label{eq:bound}
\end{equation}
where $w \equiv \hbar\bar \omega/E_F$, $\lambda$ is the (conventionally defined) dimensionless electron-phonon coupling, $\mu$ is the dimensionless electron-electron repulsion, and $\ldots$ represents other things such as the form of the electron dispersion ({\it e.g.} the ratio of second to first neighbor hopping matrix elements), 
the phonon dispersion, etc.  

Given that in typical metals, the Fermi energy is large compared to all other energies, it is reasonable to evaluate $A$ in the 
$w \to 0$ limit.  In this same limit, $\mu$ is highly renormalized downward, so that  $\mu \to 0$;  in any case, $\mu>0$ will only tend to reduce $A$. Thus, the most significant parametric dependence of $A$ concerns $\lambda$.

Various approximate theoretical treatments of this problem based on Migdal-Eliashberg (ME) theory have long served as the basis for the accepted wisdom on this subject.  Since nominally ME theory is valid so long as $w \lambda \ll 1$, it is generally believed that it can be applied even to the strong coupling regime in which $w^{-1} \gg \lambda \gg 1$.  Moreover, various approximate evaluations of the resulting self-consistency equations produce an expression for $A$ that is a monotonically increasing function of $\lambda$;  if correct, this would imply that, barring other instabilities {\it i.e.} lattice instabilities or charge-density wave (CDW) formation, the larger $\lambda$ the larger $T_c$.

We have recently shown\cite{PhysRevB.97.140501}, on the basis of exact DQMC studies of the Holstein model (defined below), that this expectation is incorrect.  Specifically,  for a given small value of $w$ we find that  the ME approximation is extremely accurate for  $\lambda$ smaller than  a characteristic value $ \lambda^\star \sim 1$, while  for $\lambda > \lambda^\star$ (even if $\lambda \ll w^{-1}$) ME theory is both quantitatively and qualitatively incorrect.  In particular, $\lambda^\star$ marks  a crossover to a strong-coupling regime characterized by bipolaron formation, growth of commensurate CDW correlations unrelated to  Fermi surface nesting, and incipient  phase separation. 

The Holstein model consists of a single electronic band, and a   non-dispersing optical phonon coupled by the most local possible interaction to the on-site electron density. In relating model calculations with experiment, it is important to distinguish ``bare parameters'' (i.e. the parameters that appear in the model, such as the bare phonon frequency, $\omega_0$ and the bare dimensionless electron-phonon coupling, $\lambda_0$) from  ``physical''  quantities, such as the actual phonon dispersion, $\omega_{\vec q}$, and the renormalized electron-phonon coupling, $\lambda$. The  tendency to phonon softening with increasing $\lambda_0$ implies that $\omega_{\vec q} < \omega_0$ and $\lambda > \lambda_0$.  Indeed, in the context of ME theory, $\lambda$ diverges upon approach to a lattice instability. 
The breakdown of ME theory we have identified is distinct from any such lattice instability, and occurs when  $\lambda$ and $\lambda_0$ are both of order 1.  The crossover at $\lambda^\star$ is associated with a complete rearrangement of the important low energy degrees of freedom.  

\begin{figure}[t!]
    \centering
    \includegraphics[width=4.5in]{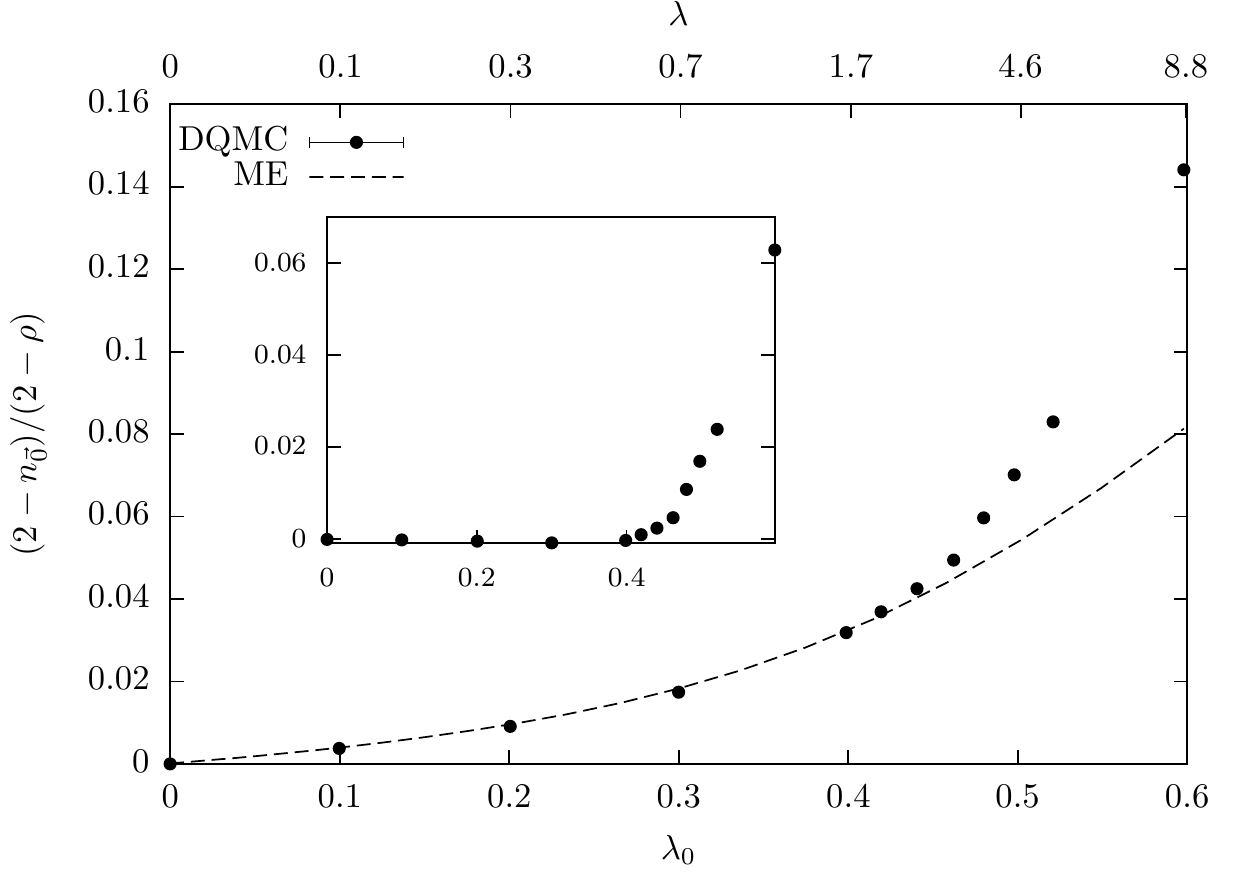}
    \caption{Occupancy of the $\vec k = \vec 0$ (band bottom) single-particle state as function of the bare dimensionless electron-phonon coupling $\lambda_0$ (lower scale) and the renormalized coupling, $\lambda$ (upper scale). (The definitions are as given in \cite{PhysRevB.97.140501}). Solid dots are from DQMC simulations described in \cite{PhysRevB.97.140501} 
    with $\hbar \omega_0/E_F = 0.1$, and the data shown are at a  temperature $T \approx E_F/30$, at which $n_{\vec 0}$ has reached its asymptotic low $T$ value. 
    The dashed line is the same quantity computed within ME theory. The inset shows the difference between the DQMC and ME results. The 
    scale on the vertical axis has been chosen such that 
    0 is the value for non-interacting electrons and 1 is the value in the polaronic limit.($\rho=0.8$ is the average electron density per site.)} 
    \label{fig:k0_occ}
\end{figure}

To quantify this crossover we show in Figure \ref{fig:k0_occ} the $T\to 0$ occupancy of the single particle state at the bottom of the band ($n_{\vec k}$ for $\vec k=\vec 0$), measured in DQMC and computed within ME theory. This state is far below the Fermi energy and hence, for non-interacting electrons 
$n_{\vec 0}=2$.  On the other hand, in the strong-coupling polaronic limit, electrons are essentially localized on a lattice site, so 
$n_{\vec  0}$ approaches the 
average electron density per site. For $0 <\lambda \ll 1$ the electronic spectrum is perturbatively rearranged and $n_{ \vec 0}$ is slightly depressed. This behavior is apparent in both the ME approximation and from the DQMC results for $\lambda < \lambda^\star$. However, for $\lambda > \lambda^\star$, 
the DQMC results show a rapid decrease of $n_{ \vec 0}$, consistent with a crossover to the polaronic limit. Moreover, even though we are not able to directly compute $T_c$ (due to the difficulty in obtaining convergence of the DQMC results at low temperatures), by a series of indirect arguments, we\cite{PhysRevB.97.140501} inferred that $T_c$ is maximal 
near the point of this crossover, $\lambda = \lambda^\star$, and decreases dramatically when $\lambda$ is either decreased {\em or increased} further.  This leads us to the conclusion that there is a well-defined maximal value  ${\rm Max}[A] \equiv A_{max} = A(0,\lambda^\star,0,\ldots)$.

It turns out that a careful numerical evaluation of the full ($\vec k$ and $\omega$ dependent) self-consistent ME equations also leads to a non-monotonic behavior of $A$, which leads to a vanishing $T_c$ for $\lambda > \lambda^\star$.  However,  in contrast to what is found in the DQMC, in the ME treatment the depression of $T_c$ for $\lambda > \lambda^\star$ is associated with the onset of a competing incommensurate CDW order. This distinction is important, since if it were only the competition with CDW order that prevented high $T_c$s, one could ``engineer'' interactions\cite{Pickett2006} that suppress CDW order so as to enhance $T_c$. 

Combining the results from the ME theory (where valid) with the DQMC results, we obtained  estimates of $A_{max}$ 
for the Holstein model on the square lattice. 
For the range of parameters we have explored, the highest value of $T_c$ we have inferred is 0.08 times the bare phonon frequency, but because significant phonon softening occurs for $\lambda \sim \lambda^\star$, this value of $T_c$ is 0.12 times the maximal renormalized phonon frequency. 
Many physically realistic generalizations of this model are possible - either by modifying the lattice structure, the electron band-structure (further neighbor hopping matrix elements), the number of phonon modes and their dispersion, and the structure of the electron-phonon coupling - all features represented by the $\ldots$ in $A$.  There is no reason that the value of $A_{max}(\ldots)$ obtained by optimizing with respect to $\lambda$, should not depend somewhat on these various features, although we have already found it to be relatively insensitive to small changes  of the band structure.  Still, it is an interesting exercise (which we are currently undertaking) to determine what microscopic features of the electron-phonon problem can 
increase $A_{max}$.  

For now, however, we will adopt an estimate of $A_{max} \approx 1/10$ as suggested by results for the simple Holstein model, and see how it compares with experiment in real materials.


\section{Experimental determination of $k_BT_c/\hbar \bar \omega$}
\begin{figure}[t!]
    \centering
    \includegraphics[width=4.in]{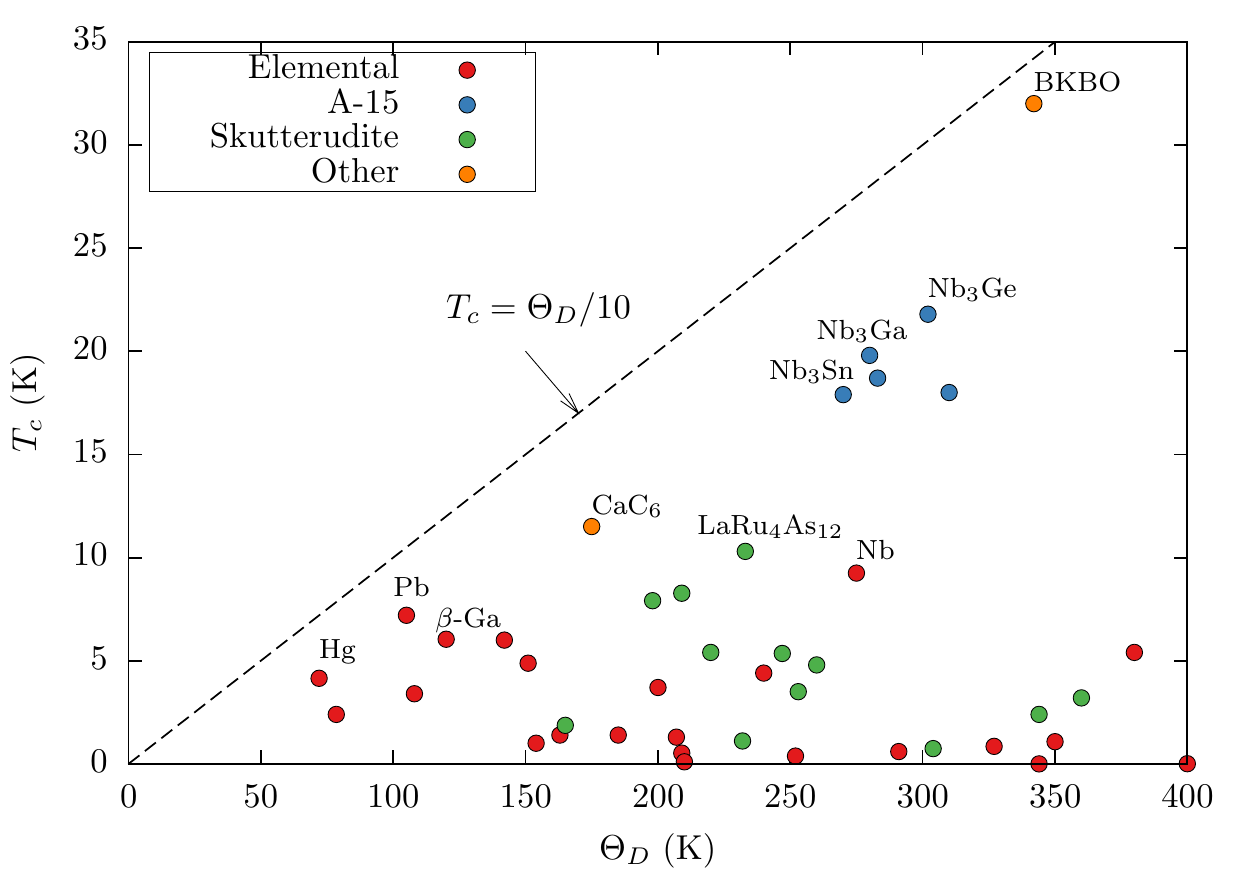}
    \caption{Measured values of $T_c$ and $\Theta_D$ for various putatively conventional crystalline superconductors. The shaded region is the proposed bound  $T_c = (0.08 - 0.12)\ \Theta_D$. Data on the various materials were obtained from the following references: elements from \cite{webb2015superconductivity,campanini2018raising}; A-15s from \cite{stewart2015superconductivity},  Skutterudites from \cite{sato2009chapter, 1742-6596-534-1-012040,juraszek2016specific}; $\mathrm{CaC}_6$ from \cite{PhysRevB.76.052511}; BKBO from \cite{PhysRevB.42.4794}. 
    In most cases estimates of $\Theta_D$ were obtained from low-temperature specific-heat measurements, in others it was estimated from {\it e.g.} low temperature resistivity. This data set is far from exhaustive and we plan to expand on it in the future.}
    \label{fig:Holstein_bound}
\end{figure}

In Fig. \ref{fig:Holstein_bound} we plot the superconducting $T_c$ vs. the Debye temperature, $\Theta_D$ for various elemental superconductors and compounds for which data is available.  In most of the data shown, $\Theta_D$ is computed on the basis of the measured low temperature lattice contribution to the specific heat and  the number of atoms per unit cell in the crystal structure, while in others it is inferred from {\it e.g.} low temperature resistivity. It thus  represents a specific, unambiguously defined (although somewhat crude) estimate of the characteristic phonon frequency $\hbar\bar \omega \sim k_B\Theta_D$.  Also shown in the figure is the proposed bound $T_c \leq \Theta_D/10$. 

Not only do we see that the bound is satisfied by all the data we have found (which is not, unfortunately, an exhaustive set), but in some cases the materials come quite close to saturating the bound, meaning that the bound may have some real significance.  Specifically, for Pb ($T_c=7.2$ K), Nb ($T_c=9.25$ K), and Hg ($T_c=4.15$ K), three elemental superconductors known for their relatively strong electron-phonon couplings, $T_c/\Theta_D$ 
takes on the values 0.069, 0.034, and 0.058, respectively.  The A-15 family of old fashion ``high temperature superconductors," Nb$_3$Sn ($T_c=17.9$ K), Nb$_3$Ga  ($T_c=19.8$ K), and Nb$_3$Ge  ($T_c=21.8$ K) have $T_c/\Theta_D$ 
equal to 0.066, 0.071, and 0.072, respectively.  

At ambient pressure, the highest temperature superconductivity of a clearly conventional sort (with $w \ll 1$) is MgB$_2$, which does not appear in the figure because it has such a high $\Theta_D = $ 884 K, and thus has $T_c/\Theta_D = 0.04$ \cite{0953-2048-14-11-201}.  This suggests that if a way can be found to increase the value of $\lambda$ in this material, it could lead to as much as a factor of 2 enhancement of $T_c$.  The highest $T_c$ of all conventional superconductors is $T_c=203$ K in H$_3$S at 155 GPa \cite{drozdov2015conventional}.  As far as we know, the Debye temperature has not been measured;  however, 
if we identify $\bar \omega$ with the largest phonon frequencies found in DFT calculations of the phonon band structure we obtain the estimate $\hbar \bar \omega=0.23$ eV \cite{PhysRevLett.114.157004}.  If we accept this theoretical value, then $k_BT_c/\hbar\bar \omega = 0.08$, {\it i.e.} it comes very close to saturating our bound. 

An especially interesting material from our perspective is $\mathrm{Ba}_{1-x}\mathrm{K}_x\mathrm{BiO}_3$ (BKBO), which has an optimal $T_c = 32 K$ ($T_c/\Theta_D \approx 0.09$) for $x \approx 0.4$ \cite{PhysRevB.42.4794}. Various features \cite{hundley1989specific,Baumert1995} of BKBO near optimal $T_c$ -- softening of an optic phonon mode, diamagnetism above $T_c$, proximity to a commensurate CDW phase -- indicate this material may be especially relevant for studying the crossover from conventional superconductivity to strong-coupling, polaronic physics.

\section{Further remarks}
The idea of bounding $T_c$ is not new.  For instance, an absolute bound for an electron-phonon mechanism $T_c < 30$K was proposed in \cite{doi:10.1063/1.2946185}.  
The loopholes in this analysis were recently summarized in \cite{PhysRevB.74.094520}.
Conversely, 
 a remarkable and highly influential analysis\cite{PhysRevB.12.905}  of ME theory suggested that 
$T_c$ grows without bound with increasing $\lambda$, and can even be larger than $\hbar\bar \omega/k_B$.  It has long been recognized that this proposal was subject to the caveat that at large $\lambda$, a system may be prone to other instabilities, which could compete with superconductivity.  It was shown\cite{PhysRevB.74.094520} that the effect of phonon softening  - still assuming the validity of ME theory - leads to a reduced prefactor in the $T_c$ expression, and thus to a bound on $T_c$ given by the bare phonon frequency, which is approached asymptotically as $\lambda \to \infty$.  Arguments for a bound within the context of ME theory have also been presented in \cite{varma2012considerations} and \cite{leavens1975least}.

While the analysis  in the present paper shares some features with these earlier studies, there are important ways in which it is different, both conceptually and practically.  Our DQMC results show that deviations from the predictions of ME theory occur even in ranges of temperatures and $\lambda$ in which no other form of order has arisen;
in this range, ME theory always over-estimates the superconducting susceptibility.  
The non-monotonic $\lambda$ dependence of $T_c$ that leads to our proposed bound is associated with a crossover at  $\lambda \sim \lambda^\star$ from a regime in which ME theory is extraordinarily accurate to a strong-coupling regime where the ME approximation breaks down entirely.  This leads to a sharp drop of $T_c$, even if the system in question has been carefully engineered to have no competing charge-ordering instabilities.  
The existence of an optimal $\lambda^\star$  independent of competing instabilities, as far as we know, is inconsistent with all analysis based on ME theory, but  not in conflict with any experimental observation.

Recently, it has been found that superconductivity in SrTiO$_3$ persists to such low electron densities that the Fermi energy is less than the typical phonon frequency, {\it i.e.} into a regime in which $w \gg 1$.  
In this limit, it is far from clear that the effects of a bare repulsion, $\mu>0$, can still be neglected. 
However, as a problem in model physics, it is possible to ask whether a more general bound exists on $A(w,\lambda,\mu=0,\ldots)$.  In the large $w$ limit, it is possible to integrate out the phonons to obtain an instantaneous attractive interaction with a magnitude proportional to $\lambda$.  In particular, the Holstein model in the limit $w\to\infty$ maps onto the negative $U$ Hubbard model with $U \sim - E_F \lambda$.  It is well known that this model has a optimal $T_c$ at an intermediate value of $|U| \sim E_F$ \cite{PhysRevB.69.184501}. 
Thus, at least in the artificial limit $\mu=0$, there exists a more general bound of the form $A(w,\lambda,\mu=0, \ldots) \leq  A(w,\lambda^\star(w),\mu=0,\ldots)$ where
$A(w,\lambda^\star(w),\mu=0,\ldots) \sim 1/w$ as $w \to \infty$.

There is another class of bounds on $T_c$ that can be inferred in a different manner;  rather than taking normal state data, one can start with measured properties of the superconducting groundstate, from which one can attempt to bound the actual $T_c$.  Ideally, these bounds should apply to both conventional and unconventional superconductors.

For a simple BCS s-wave superconductor in the weak coupling limit, $T_c = B \Delta_0$ where $\Delta_0$  is the zero-temperature gap and $B = e^\gamma / \pi \approx 0.567$.  Strong coupling effects, even in the context of BCS mean-field theory, have a tendency to decrease the value of $B$, and certainly fluctuational effects beyond mean-field theory will likewise decrease the value of $B$.  Thus, taking into account the fact that the gap function can vary along the Fermi surface, one would generally expect that
\begin{equation}
    k_B T_c \leq B |\Delta_{max}|
\end{equation}
where  (with the case of unconventional superconductors in mind) $\Delta_{max}$ is the largest value of the $T=0$ gap on the Fermi surface.  As far as we know, this inequality is satisfied (and often very nearly saturated) by all known crystalline superconductors.  (In the presence of disorder it is, of course, possible to have gapless superconductors.)

In a similar vein, a 
bound was proposed in Ref. \cite{PhysRevLett.74.3253} based on the measured value of the zero temperature superfluid stiffness, $\rho_s(T=0)/m^\star$ (or equivalently from the zero temperature value of the London penetration depth)
\begin{equation}
    k_BT_c \leq C\ \frac{\hbar^2 \rho_s(0) a}{2m^\star}
\end{equation}
where in a layered (quasi-2D) superconductor, $a$ is the inter-layer separation while in a 3D superconductor, $a = \sqrt{\pi}\xi_0$ where $\xi_0$ is the zero-temperature coherence length, and $C\approx 2.2$. ($2.2$ is the ratio of $T_c$ to the zero temperature phase stiffness of the 3D XY model on a cubic lattice.)  Most superconductors satisfy this inequality but come nowhere near saturating it.  However, certain high temperature superconductors, especially hole-doped cuprates, come within a factor of two or three of saturating this inequality.   

\section*{Acknowledgements}
We thank T. Geballe and M. Beasley for bringing up the case of BKBO and A. Chubukov for useful discussions.  IE was supported by the U.S. Department of Energy, Office of Basic Energy Sciences, Division of Materials Sciences and Engineering, under Contract No. DE-AC02-76SF00515. SAK and IE were supported, in part, by NSF grant \# DMR-1608055 at Stanford.  DJS acknowledges support from SciDAC, U.S. Department of Energy,  Advanced Scientific Computing Research and Basic Energy Sciences, Division of Materials Sciences and Engineering.

\providecommand{\href}[2]{#2}\begingroup\raggedright\endgroup

\end{document}